\documentclass{article}

\usepackage{graphicx}
\usepackage{psfig}
\usepackage{epsfig}
\usepackage[round]{natbib}

\title{
Statistical modelling of tropical cyclone tracks: modelling
cyclone lysis
}

\begin{document}

\author{Tim Hall, GISS\footnote{\emph{Correspondence address}: Email: \texttt{tmh1@columbia.edu}}\\and\\
Stephen Jewson\\}

\maketitle

\begin{abstract}
We describe results from the fifth stage of a project to build a
statistical model of tropical cyclone tracks. The previous stages considered
genesis and the shape of tracks.
We now consider in more detail how to represent the lysis (death) of tropical cyclones.
Improving the lysis model turns out to bring a significant improvement to
the track model overall.
\end{abstract}

\section{Introduction}

The impact of tropical cyclones on communities in various parts of
the world motivates the development of accurate methods for
estimating tropical cyclone risks. One set of methods that have
been used for estimating these risks are the so-called basin-wide tropical
cyclone models, examples of which have been described
in~\citet{fujii}, \citet{drayton00}, \citet{vickery00} and \citet{emanuel05}.
These models attempt to represent the entire life-cycle of tropical cyclones
statistically, and thus allow the generation of an unlimited number of
random but realistic cyclones that can be used for risk estimation.
This article describes the latest results from a
research effort aimed at developing new and more accurate basin-wide tropical cyclone
models. Our previous articles on this subject have described models for
mean tropical cyclone tracks~\citep{hallj05a}, for the variance of
fluctuations around the mean~\citep{hallj05b}, for the autocorrelation of
the fluctuations around the mean~\citep{hallj05c}, and for genesis~\citep{hallj05d}.
\citet{hallj05c} also included a very simple representation of the death, or \emph{lysis},
of tropical cyclones.
We now describe a model that gives a more realistic representation of lysis, by
modelling the probability of lysis as a function of location.
We find that this significantly improves the performance of the track simulation overall.

\section{Data}

We continue with our focus on tropical cyclones in
the Atlantic basin,
and the data we use is the `official' National Hurricane Center
track data set, known as HURDAT~\citep{hurdat}. We only use data from 1950, since this is the only
data that we consider to be sufficiently reliable.
HURDAT data from 1950 to 2003 contains 524 tropical cyclones, and
these are the data that we use.

\section{Methods}

In~\citet{hallj05c} we show simulations of Atlantic hurricane tracks, and make a number
of comparisons between the simulated tracks and the historical observations.
The simulated tracks are somewhat realistic, but a
detailed comparison with observations indicates various systematic
errors. We are gradually testing various possible improvements to
the model in order to reduce these errors to a more acceptable
level. One process that was only represented in a very rudimentary
fashion in~\citet{hallj05c} is the lysis of
hurricanes: simulated hurricanes originate from
historical genesis points, and each is given a lifetime equal to
the lifetime of the historical hurricane that originated from the same point. This
gives a realistic distribution of hurricane lifetimes overall (by
definition) but also leads to some physical inconsistencies.
For instance, some simulated hurricane tracks wander
into regions where real hurricanes would die, and too many
hurricanes die in regions where real hurricanes only die
infrequently. This, in turn, affects the distribution of the density
of hurricane tracks in different regions, and would ultimately affect
estimates of risk.

To remedy these problems we now describe a more complex lysis model.
The model is very similar to the models we use for genesis and
track dynamics in the papers cited above in that it is a
non-parametric model based on spatial averaging of the behaviour of the
observations.
The optimum averaging
length-scale is determined using cross-validation, with `optimum' defined
by maximising the out-of-sample likelihood. At each time step, every simulated hurricane has
a probability $p$ of lysis. This probability varies in space,
and is determined as follows:

\begin{itemize}
    \item A two dimensional Gaussian window, with lengthscale $L$, is constructed around the current
    location of a simulated hurricane.
    We then note that every historical storm point in the basin is either the last point of a hurricane track,
    or not.
    \item $p$ is estimated as the sum of the Gaussian weights over the last points divided by the sum
    of the Gaussian weights over all points.
\end{itemize}

The lengthscale $L$ is determined from historical data using a
maximum out-of-sample likelihood jack-knife cross-validation
scheme as follows:

\begin{itemize}

    \item For a given value of $L$, for each year, and for each observed hurricane point
    in that year, we calculate the value of $p$ as above, but only using data from all \emph{other} years.

    \item The likelihood for the actual outcome for that hurricane point is then calculated.
    This likelihood is either $p$ if the hurricane point is the last for that track,
    or $1-p$ otherwise.

    \item The likelihood for each year is calculated as the sum of the likelihoods for all points
    in that year.

    \item The lengthscale $L$ is varied to find the maximum of the likelihood.

\end{itemize}

There are various benefits from using a non-parametric scheme based on the methods
of cross-validation and out-of-sample likelihood, such
as the avoidance of overfitting and the optimisation of out-of-sample
performance of the model. These benefits are discussed in more detail
in our previous articles.

\section{Results}

We now show some results from the model described above.

\begin{figure}[!hb]
  \begin{center}
    \scalebox{0.6}{\includegraphics{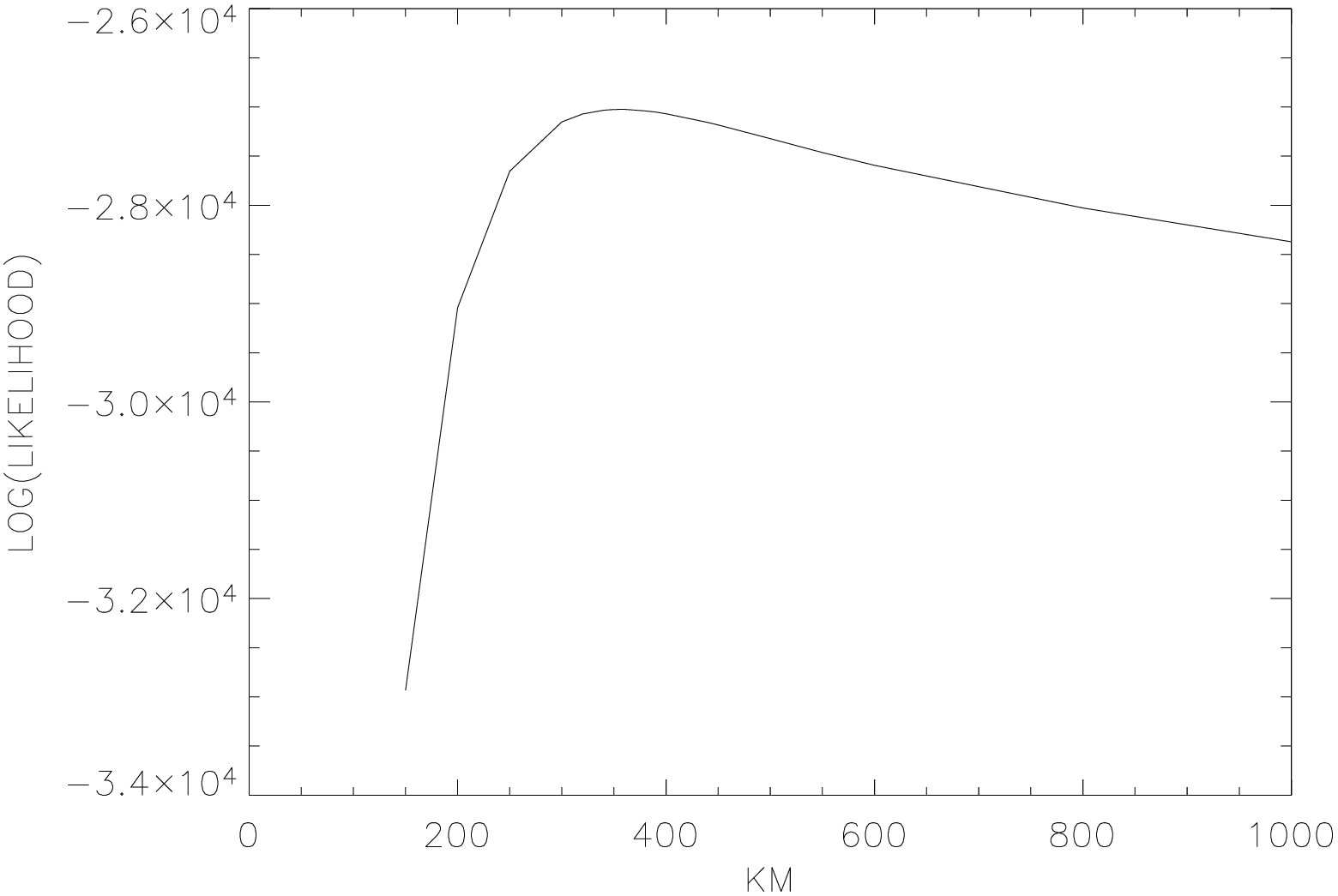}}
  \end{center}
    \caption{
The variation of the out-of-sample likelihood with the averaging length-scale for the lysis
model described in the text.
     }
     \label{f01}
\end{figure}

First, in figure~\ref{f01}, we show the out-of-sample likelihood
as a function of the lengthscale $L$. There is a very clear
maximum at 360km (accurate to within a tolerance of 10km).
The maximum represents a balance between two competing effects:
(1) decreasing L increases random sampling errors,
as fewer lysis events receive significant weight;
and (2) increasing L results in averaging of real spatial structure in the lysis distribution.

\begin{figure}[!hb]
  \begin{center}
    \scalebox{0.8}{\includegraphics{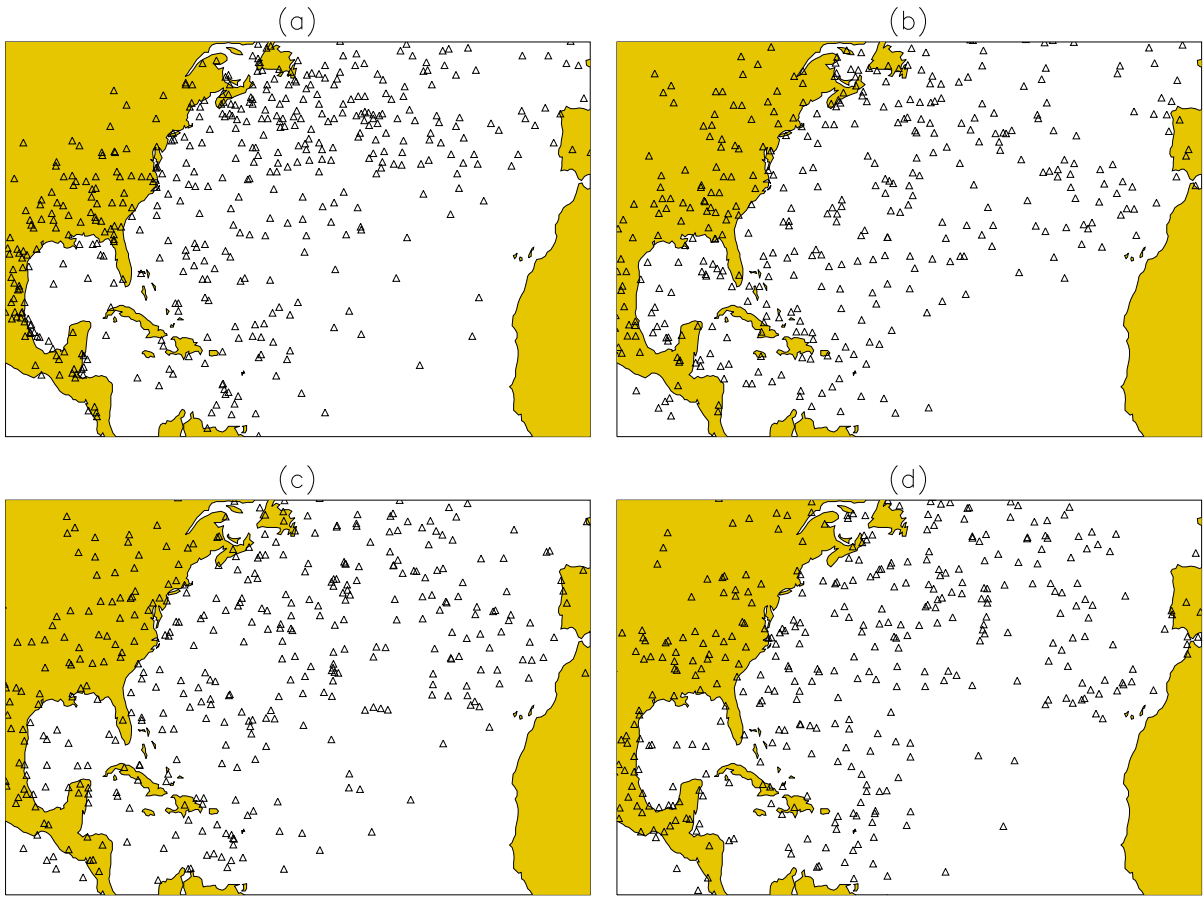}}
  \end{center}
    \caption{
The observed locations of hurricane lysis for the 524 storms from 1950 to 2003 (panel (a)),
and three simulations of hurricane lysis, each for 524 storms, for the model described in the text.
     }
     \label{f02}
\end{figure}

In figure~\ref{f02} we show the locations of historical hurricane
lysis (panel (a)) and three realisations of simulations of lysis from the
track model of~\citet{hallj05c}, but with the new lysis scheme.
The new lysis model is
reasonable, in that the overall distribution of locations of
observed hurricane lysis is captured, but it is not perfect.
There are too few deaths off the coast of Newfoundland
and too many in the region North of the Canary Islands.
We think that these discrepancies are caused by the track model,
rather than lysis directly.
A good lysis model is necessary, but not sufficient, for a realistic lysis distribution,
because the lysis distribution and the tracks are coupled.
For reasons we have yet to determine there is a weak tendency
for our simulated tracks to curve eastward too far south.
The new lysis model has greatly improved this problem compared to Hall and Jewson (2005c),
but a small discrepancy remains.
A few too many simulated storms move into the eastern subtropical
Atlantic where historical storms, and, thus, historical lysis events, are rare.
The lysis model must rely on remote lysis events to ultimately kill these storms.
This is an advantage, as it prevents the simulated storms from wandering even further a field.
However, it also generates lysis events where none occur historically.

\begin{figure}[!hb]
  \begin{center}
    \scalebox{0.8}{\includegraphics{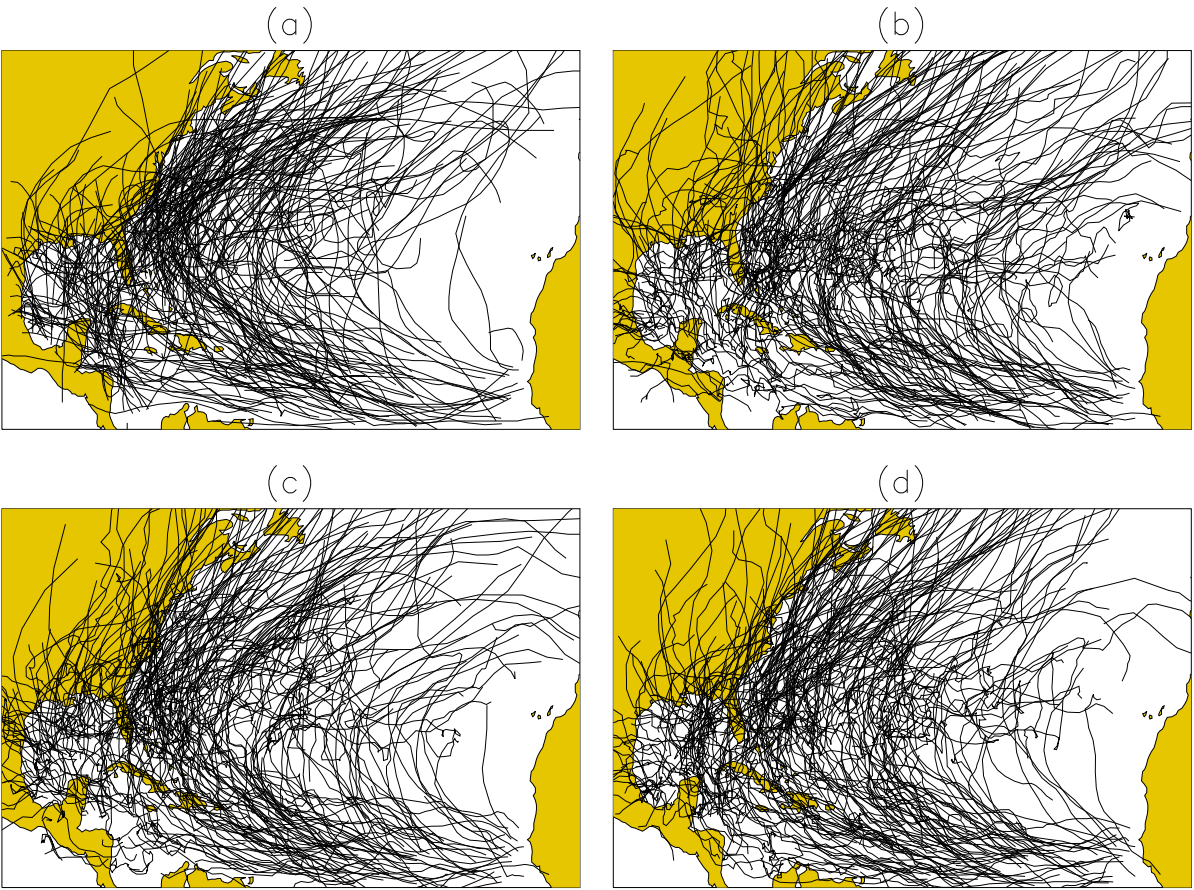}}
  \end{center}
    \caption{
Observed (panel (a)) and three realisations of simulated hurricane tracks from the model described in the text.
     }
     \label{f03}
\end{figure}

Figure~\ref{f03} shows tracks simulated from the track model. This figure should be compared with
figure 14 in~\citet{hallj05c}. In comparison with that figure, we see that (a)
the spurious point of track convergence in the East Atlantic has
disappeared and (b) the high density of tracks off the East Coast
of the US has been simulated more realistically.
However, as noted above, there is still a weak tendency for storms
to bend too far eastward in the subtropical Atlantic.

\begin{figure}[!hb]
  \begin{center}
    \scalebox{0.8}{\includegraphics{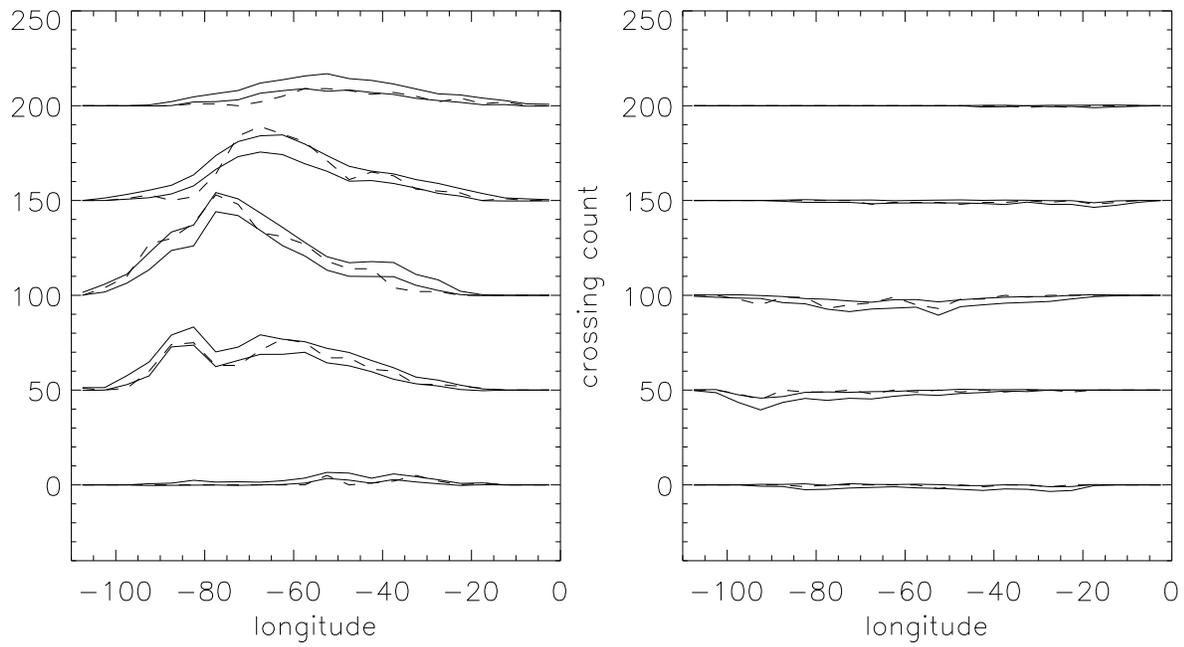}}
  \end{center}
    \caption{
  The number of tracks from observations and from simulations that cross certain lines
  of latitude (equally spaced from 10N to 50N, from bottom to top), in a northward
  direction (left panel) and in a southward direction (right panel).
     }
     \label{f04}
\end{figure}

\begin{figure}[!hb]
  \begin{center}
    \scalebox{0.8}{\includegraphics{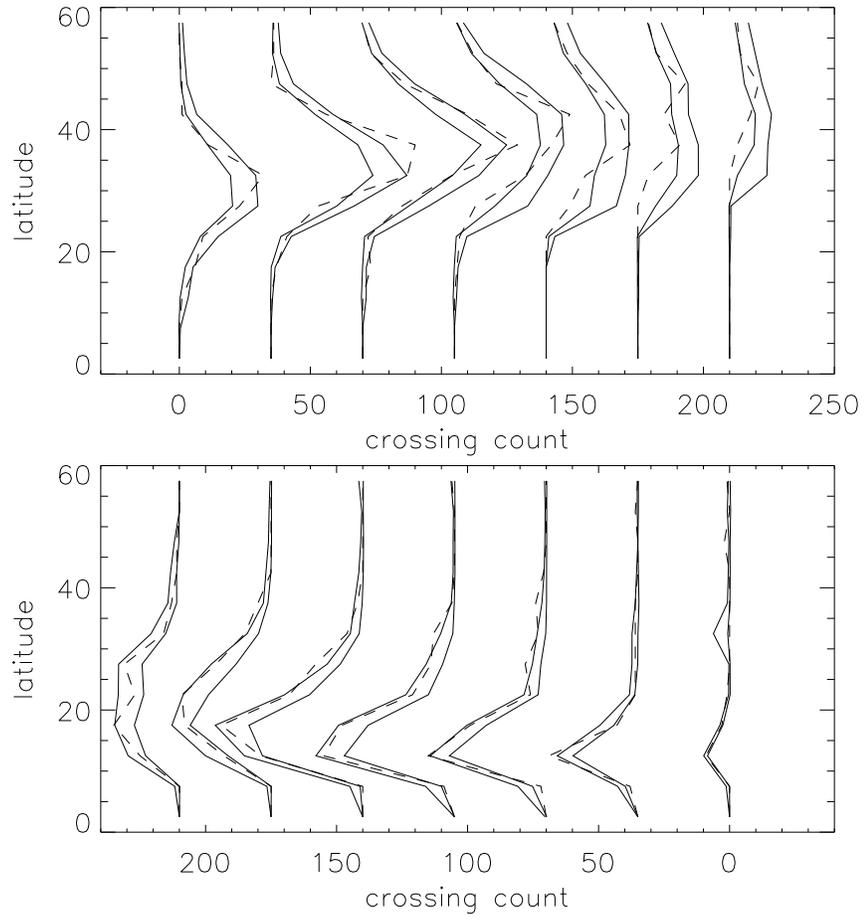}}
  \end{center}
    \caption{
  The number of tracks from observations and from simulations that cross certain lines
  of longitude (equally spaced from 80W to 20W, from left to right), in a eastward
  direction (panel (a)) and in a westward direction (panel (b)).
     }
     \label{f05}
\end{figure}

We now consider a more quantitative evaluation of the performance
of the track model with the new lysis process.
Figure~\ref{f04} is a new version of figure 17 from~\citet{hallj05c}, and
figure~\ref{f05} is a new version of figure 18. They are created by
simulating 20 realisations from the track model, each with 524
hurricanes. These 20 realisations define a distribution for the
numbers of tracks crossing lines of latitude and longitude in a 54 year track set, and
these distributions can be compared with the observed numbers of
tracks crossing these lines. One would hope that the observations
would lie within the distribution created from the statistical
model. In figure~\ref{f04}, which shows numbers of tracks crossing lines
of latitude, we see close consistency between the observations (dotted line)
and the distribution of simulated tracks (illustrated using solid lines
at plus and minus one standard deviation).
The results are much better than those from the previous model, and the systematic error
of having too few storms at certain latitudes has been fixed (it seems that the problem was
that too many storms were dying too early in their life, and were not reaching the
region of high storm density off the east coast of the US).
In figure~\ref{f05}, which shows numbers of tracks crossing lines of longitude,
the model results at low latitudes are consistent with the observations. The results at higher
latitudes (in the westerlies) are good, but there are still some slight discrepancies
in the far east of the domain. Again, however, the major systematic errors in the
previous model have disappeared.

\begin{figure}[!hb]
  \begin{center}
    \scalebox{0.8}{\includegraphics{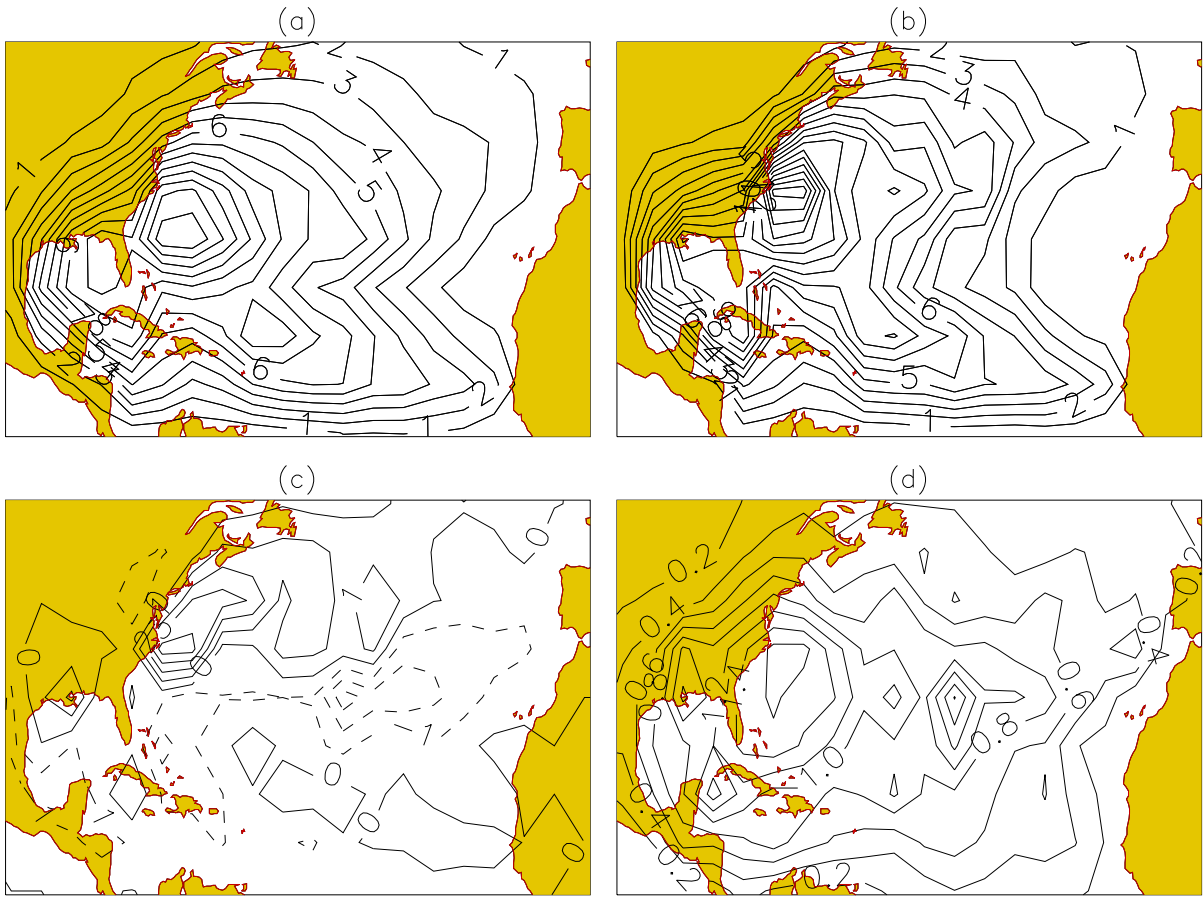}}
  \end{center}
    \caption{
  Track densities for model and observations.
  Panel (a) shows the track density for the model, averaged over 20 realisations
  of 524 storms. Panel (b) shows the track density for observations, for 524 storms.
  Panel (c) shows the difference of these densities, and panel (d) shows the
  standard deviation of the density from the model (across the 20 simulations of 524 storms).
     }
     \label{f06}
\end{figure}

Figure~\ref{f06} is a repeat of figure 20 from the earlier paper, but now for
the new track model. The first panel shows the density of
hurricane points from the simulations (based on the ensemble of 20
realisations used in the previous figures), while panel (b) shows the
same for observations. Panel (c) shows the difference between the two,
and panel (d) shows the standard deviation of the ensemble.
The track model now simulates the density of tracks consistently
with the observations almost everywhere, except in a very small
region off the coast of Georgia and the Carolinas. In this region
the observations show a very high density of tracks. The model
also simulates a high density of tracks in this region, but the peak
density is lower and the location of the maximum is
shifted slightly to the south. These differences appear to be
significant.

\subsection{Landfalling rates}

\begin{figure}[!hb]
  \begin{center}
    \scalebox{0.8}{\includegraphics{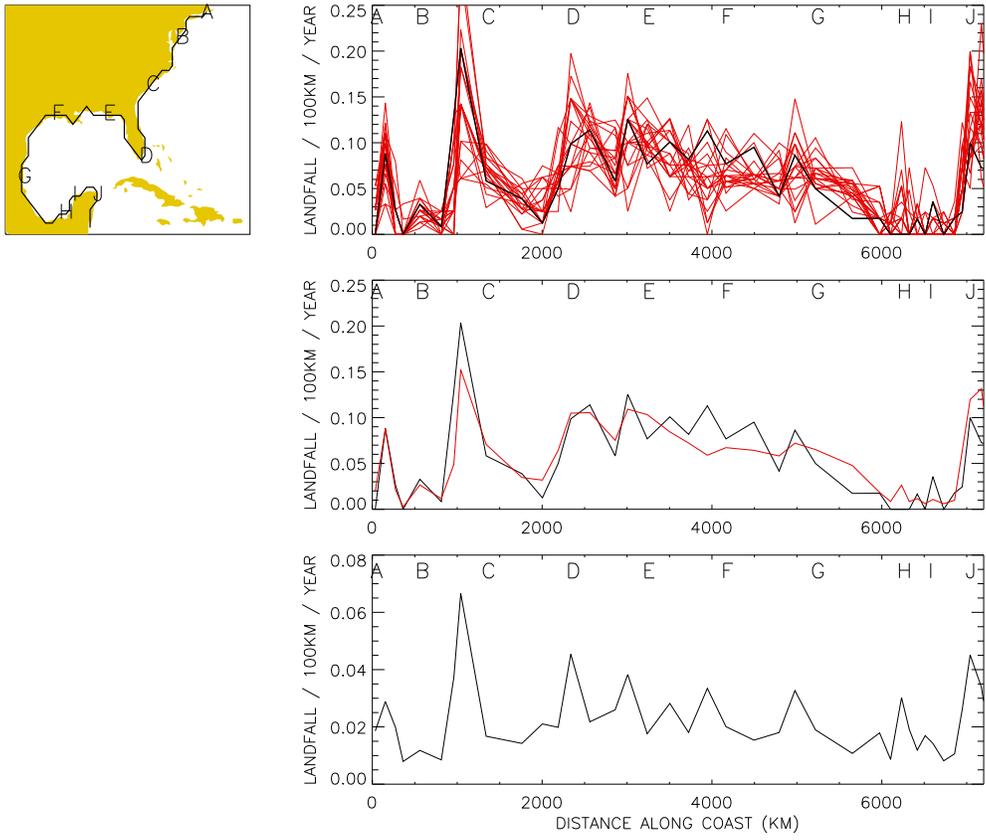}}
  \end{center}
    \caption{
The left hand panel shows a model for the coastline of North and Central America, consisting
of 39 segments. The right hand panels show various diagnostics for the number of hurricanes
crossing each of these segments in the observations and in the model.
The top panel shows 54 years of observations (black line) and 20 realisations of 54 years from the model (red lines).
The middle panel shows the observations and the mean of the model realisations.
The lower panel shows the standard deviation of the model realisations.
     }
     \label{f07}
\end{figure}

\begin{figure}[!hb]
  \begin{center}
    \scalebox{0.8}{\includegraphics{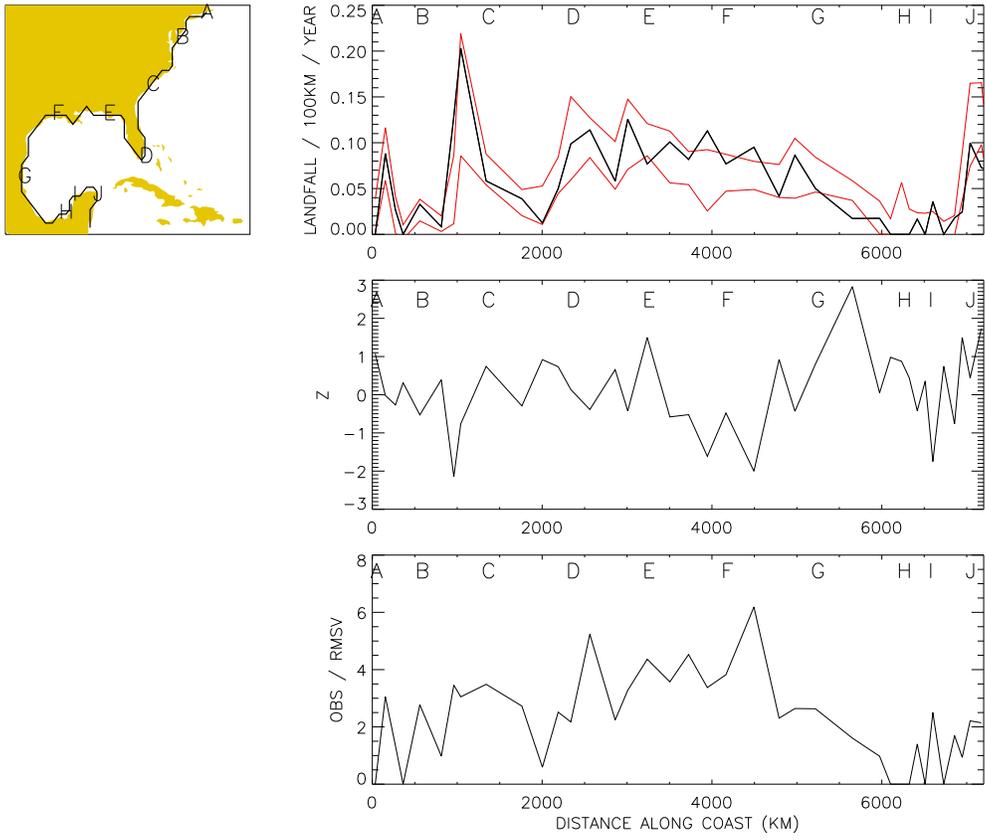}}
  \end{center}
    \caption{
As for figure~\ref{f08}, but now panel (a) shows the observations, with lines for plus and minus one standard
deviation from the model, panel (b) shows the difference between model and observations normalised using the standard deviation from the
model and panel (c) shows the observations divided by the standard deviation from the model.
     }
     \label{f08}
\end{figure}

In figures~\ref{f07} and~\ref{f08} we show various aspects of a diagnostic designed
to analyse the extent to which our model simulates
the correct number of tracks making landfall in different regions.
We use a simple representation of the North and Central American
eastern coastline, as shown in the map in the left panel in each figure. This modelled
coastline consists of 39 straight line segments of different lengths
(using different length segments gives a better fit than using identical length segments).
In the right hand panels of both figures we straighten this coastline out, with the
horizontal axis running from North to South, and with 10
reference points marked on the map and the graph to help understand the
correspondence between the two.

In the top right hand panel of figure~\ref{f07} the black line shows
the observed rate at which hurricanes crossed each of the 39 straight line segments,
in numbers of hurricanes per 100km per year, based on the 54 years from 1950 to 2003.
There are very large fluctuations
in the crossing rate depending on location on the coast. The highest rates
from this analysis are at Cape Hatteras, followed by some parts of Florida.
It is clear from the high level of variability in the rates from segment to segment
that the crossing rate at each location depends in a sensitive way on the
geometry of the coastline and the geometry and variability of the hurricane tracks.
This has implications for the extent to which these results can be considered general:
if we repeated this analysis with a slightly different approximation for the coastline
then they might look very different. It is therefore only reasonable to compare crossing rates
between model and observations when both are calculated for exactly the same coastline (as we will do below).
Our crossing rates should not, however, be compared with results from the
same analysis that use different models for the coastline.
For instance, the results that are shown on page 172 of~\citet{diaz} were generated using a different
approximation to the coastline. As a result, although the
pattern of landfall rates displays similarly located minima and maxima
the relative magnitudes of these extrema are different,
and the maximum landfall rates in that analysis are in Florida rather than near Cape Hatteras.

The red lines in the top right panel of figure~\ref{f07} show the crossing rates derived from simulations from
the model. Each line corresponds to one of the 54 year realisations described above, and there are 20 lines
corresponding to the 20 realisations.
We see considerable variability between the different realisations: this suggests that
the historical observed crossing rates themselves could have been rather different, and are
very affected by randomness. For instance, some of the realisations have the highest crossing
rates in Florida rather than Cape Hatteras, and there are many locations where the
crossing rates vary by a factor of 2 or more between different 54 year realisations.
We will investigate this latter point further below.

The middle panel of figure~\ref{f07} shows the mean crossing rates for the
simulations from all 20 realisations together.
The agreement between the model and the observations
in this and the top panel
 suggests that various
features in the observations are real, rather than statistical artefacts.
These include the deep minimum of rates between points C and D, on the North East coast of Florida.

The lower right panel of figure~\ref{f07} shows the standard deviation across the 20 realisations.
There are considerable variations in this standard deviation with location, with a pattern
that resembles the mean. These variations show that not only is the mean crossing rate affected by the geometry
of the coastline and the tracks, but also the variability among different 54 year samples.

Figure~\ref{f08} shows three more ways in which we can compare the observed and simulated landfalling rates.
First, the top right hand panel shows the observed
landfall rates, with the plus and minus one standard deviation range for the model, derived
from the 20 realisations.
Second, the middle panel shows the difference between
the mean model rate and the observations, divided by the ensemble standard deviation.
Considering these two diagnostics we see that
there are two places in particular where the observations may be inconsistent with the
simulations: between B and C, and between G and H. Does this indicate flaws in the model?
In fact it seems likely that the difference between B and C is related to the errors in the simulation of
the high density of tracks off the coast of the Carolinas, as illustrated in figure~\ref{f06}
and discussed above. This is because the location of the difference in coastal rates is directly `downstream',
in a track sense, from the high density region.
The different between G and H is smaller, although more significant, is less easy to understand.

The lower right hand panel in figure~\ref{f08} shows the ratio of the observed crossing rates to the
model ensemble standard deviation. This gives an indication of where the observed rates are most
affected by randomness: high values indicate that the role of randomness is small and that the observed
rates are presumably a good estimator of the (unknown) real rates, while smaller values indicate that
there is a large effect of randomness. In general, this ratio follows the observed rates. In other words,
it is more or less the case that in regions where there are more hurricanes the landfalling rates
can be estimated more accurately than in regions where there are fewer hurricanes (as one might expect).
This is of some relevance to the insurance industry, as follows. One would expect property insurance rates
to be lower in regions where landfalling rates are observed to be lower on average.
However, one would also expect insurance rates to depend, in part, on the uncertainty in the estimation
of the hurricane rates. Some of the regions where the observed rates are low, but the uncertainty is high (i.e. there
is the possibility that the observed rates are only low by good fortune over the last 54 years),
should, perhaps, have higher insurance rates than one might otherwise expect.

\section{Conclusions}

We have described the latest version of our hurricane track model.
Relative to the previous version, we have included a more
realistic lysis process. Ultimately, we hope to model fluctuations
in the pressure of hurricanes as well as the tracks, and a good
pressure model would take care of lysis automatically. However,
pressure modelling is not simple, and at this point we have
created a purely empirical model for lysis based on the observed
distribution.

The new lysis model has
significantly improved the simulation of hurricane tracks, has removed
the major systematic errors in our model, and has
partly solved the problem of how to simulate the high density of
tracks off the US East Coast. A few smaller systematic errors remain, such as
the precise location and magnitude of the peak density of tracks in this region.

There is still more work to do, however. We have not addressed the
question of the non-normality of the residuals in the track model,
as shown in figure 11 in~\citet{hallj05c}. Presumably better modelling of the
residuals would lead to further improvements in the large-scale
behaviour of the tracks. And our model is also entirely non-seasonal
at this point, even though the
genesis of hurricanes, and possibly also the track dynamics and lysis,
are definitely not constant across the hurricane
season. If one is interested in the seasonal variation of risk
then modelling these seasonal variations is imperative. And
modelling seasonal variations correctly \emph{may} also help in
the modelling of overall annual risk.

\newpage
\bibliography{../bib/jewson}


\newpage

\newpage

\newpage

\newpage

\newpage

\newpage

\newpage

\newpage


\end{document}